\documentclass[11pt]{article}

\usepackage[letterpaper,margin=1in]{geometry}
\usepackage[T1]{fontenc}
\usepackage[utf8]{inputenc}
\usepackage{lmodern}
\usepackage{amsmath,amsfonts,amssymb}
\usepackage{array}
\usepackage{booktabs}
\usepackage{tabularx}
\usepackage{textcomp}
\usepackage{url}
\usepackage{graphicx}
\usepackage{cite}
\usepackage{xcolor}
\usepackage{listings}
\usepackage{enumitem}
\usepackage{microtype}
\usepackage{placeins}
\usepackage{float}
\usepackage[hidelinks,breaklinks=true,bookmarks=false]{hyperref}

\hypersetup{
  pdftitle={Governing AI-Assisted Security Operations: A Design Science Framework for Operational Decision Support},
  pdfauthor={Elyson A. De La Cruz; Rishikesh Sahay; Md Rasel Al Mamun},
  pdfsubject={arXiv preprint},
  pdfkeywords={AI governance, cybersecurity management, design science research, engineering management, operational decision support, security operations center, technology adoption}
}

\newcolumntype{Y}{>{\raggedright\arraybackslash}X}
\newcolumntype{L}[1]{>{\raggedright\arraybackslash}p{#1}}
\setlist[itemize]{leftmargin=*,noitemsep,topsep=2pt}
\setlist[enumerate]{leftmargin=*,noitemsep,topsep=2pt}
\definecolor{codegray}{rgb}{0.96,0.96,0.96}
\definecolor{codeborder}{rgb}{0.70,0.70,0.70}
\definecolor{darkblue}{rgb}{0.0,0.0,0.45}
\definecolor{codestring}{rgb}{0.35,0.10,0.10}

\lstdefinelanguage{KQL}{
  morekeywords={let,where,project,extend,summarize,join,kind,union,order,by,top,ago,todynamic,tostring,dynamic,has,has_any,in,in~,matches,regex,make_set,count,dcount,countif,case,iff,on,limit,distinct,mv-expand,extract_all,between},
  sensitive=false,
  alsoletter={-},
  morecomment=[l]{//},
  morestring=[b]",
}
\lstdefinestyle{kqlcompact}{
  basicstyle=\ttfamily\footnotesize,
  backgroundcolor=\color{codegray},
  breaklines=true,
  frame=none,
  rulecolor=\color{codeborder},
  framerule=0.3pt,
  columns=fullflexible,
  keepspaces=true,
  showstringspaces=false,
  keywordstyle=\color{darkblue}\bfseries,
  stringstyle=\color{codestring},
  xleftmargin=0.4em,
  framexleftmargin=0.4em,
  aboveskip=0.45em,
  belowskip=0.45em,
  captionpos=b
}
\title{Governing AI-Assisted Security Operations: A Design Science Framework for Operational Decision Support}
\author{%
Elyson A. De La Cruz\\
\small Department of Artificial Intelligence, University of the Cumberlands, Williamsburg, KY, USA\\
\small \href{mailto:elyson.delacruz@ucumberlands.edu}{elyson.delacruz@ucumberlands.edu}; ORCID: \href{https://orcid.org/0009-0006-5599-506X}{0009-0006-5599-506X}
\and
Rishikesh Sahay\\
\small Department of Management Information Systems, University of Illinois Springfield, Springfield, IL, USA\\
\small \href{mailto:rsaha@uis.edu}{rsaha@uis.edu}; ORCID: \href{https://orcid.org/0000-0002-0380-8289}{0000-0002-0380-8289}
\and
Md Rasel Al Mamun\\
\small Department of Management Information Systems, University of Illinois Springfield, Springfield, IL, USA\\
\small \href{mailto:mmamu@uis.edu}{mmamu@uis.edu}; ORCID: \href{https://orcid.org/0000-0002-2710-6599}{0000-0002-2710-6599}
}
\date{May 10, 2026}

\begin{document}
\maketitle
\begin{abstract}
Engineering managers increasingly must decide how to introduce generative artificial intelligence (AI), retrieval-augmented generation, and coding agents into high-risk operational functions without weakening accountability, privacy, cost discipline, or auditability. The central message of this study is that AI-assisted operational decision support should be managed as a governed engineering capability before it is scaled as automation. Security operations centers (SOCs) provide a suitable setting because they combine privileged telemetry, specialist expertise, software repositories, cloud services, and evidence-sensitive decisions. This study uses Kusto Query Language (KQL) and Microsoft Azure security capabilities as a bounded technical instantiation of that broader engineering management problem. KQL is read-only in ordinary query use, but read-only does not mean risk-free: AI-assisted queries can still create privacy, cost, performance, schema-validity, and decision-quality risks through broad scans, sensitive-field exposure, stale intelligence, and misleading interpretations. Using design science research, the study develops a governed AI query-broker artifact that separates AI planning from operational execution through schema-grounded retrieval, approved templates, policy validation, read-only adapters, normalized outputs, auditable agent traces, and engineering review board gates. The contribution is not a new KQL technique, security product, or detection algorithm. Rather, the study contributes a management framework for governing AI-assisted operational decision support in high-risk digital infrastructure by specifying design propositions, role accountability, maturity stages, quality gates, evaluation criteria, and evidence boundaries.
\end{abstract}

\vspace{0.5em}
\vspace{0.75em}

\section*{Managerial Relevance Statement}
Security executives and engineering managers face a practical dilemma: AI can accelerate operational analysis and detection engineering, but ungoverned AI use can create unacceptable risk in privileged telemetry environments. The managerial value of this study is a decision framework for moving AI assistance from informal experimentation to governed operational capability. KQL is used only as the illustrative technical setting: it is read-only in ordinary query use, but broad scans, sensitive-field exposure, cost overruns, invalid schemas, and misleading interpretations can still harm operations and decision quality. The proposed broker converts AI assistance into a governed capability by requiring schema grounding, approved templates, read-only execution, time and row limits, human approval for sensitive hunts, auditable traces, and engineering review board gates. Managers can use the framework to assign operating roles, define maturity stages, establish CI/CD quality gates, monitor cost and reliability, and decide when a proof of concept is ready for shadow-mode or production use. Its intended impact is safer, more auditable, and more scalable AI adoption in high-risk engineering work.

\section{Introduction}
Engineering managers increasingly face a new adoption problem: generative artificial intelligence (AI), retrieval-augmented generation, and coding agents can accelerate technical work, but they also blur accountability when they propose actions, generate code-like artifacts, summarize evidence, or invoke tools inside high-risk operating environments. Security operations centers (SOCs) are a particularly useful setting for this problem because they combine cloud platforms, software repositories, telemetry pipelines, identity systems, threat-intelligence feeds, and human judgment in evidence-sensitive workflows. Incident-response guidance emphasizes preparation, detection, analysis, communication, recovery, and continuous improvement rather than detection tooling alone \cite{nist_sp80061r3}. The resulting challenge is not simply whether an AI system can generate a useful technical artifact. It is whether an organization can govern how that artifact is proposed, validated, approved, executed, audited, and improved.

This study uses Kusto Query Language (KQL) and Microsoft Azure security capabilities as the empirical and technical instantiation because they make the governance problem observable in a real enterprise-scale environment. Microsoft describes Kusto queries as read-only requests that process data and return results, while distinguishing them from management commands \cite{ms_kql_overview}. Microsoft Graph provides the \texttt{security/runHuntingQuery} method for advanced hunting in Microsoft Defender XDR and identifies \texttt{ThreatHunting.Read.All} as the relevant permission for the Graph advanced-hunting API \cite{ms_graph_hunting}. Defender XDR advanced hunting and Microsoft Sentinel support cross-source investigation across endpoint, identity, email, cloud, and SIEM data sources \cite{ms_defender_hunting}. Microsoft Sentinel analytics, Sentinel data lake, and Azure Data Explorer add additional query, retention, and analytics contexts \cite{ms_logs_api,ms_sentinel_lake,ms_kusto_api}. These characteristics make the setting suitable for engineering management research because it exposes real choices about permissions, schemas, cloud cost, retention, auditability, role assignment, and evidence quality.

KQL's read-only nature is important but insufficient as a governance argument. It lowers the blast radius compared with response automation that isolates devices, disables accounts, deletes messages, or changes cloud configurations. It does not eliminate operational risk. AI-generated read-only queries can still reference unavailable tables, omit required time filters, perform excessive scans, expose sensitive fields, increase cloud or query cost, misuse weak threat intelligence, or produce misleading interpretations. Telemetry fields such as command lines, URLs, filenames, and email subjects may also contain attacker-controlled text that should be treated as evidence rather than instructions. The study therefore uses KQL as a bounded technical instantiation of a broader management problem: governing AI-assisted operational decision support in high-risk digital infrastructure.

The study addresses the following research question: \emph{How can engineering managers govern AI-assisted operational decision support in security operations while preserving technical validity, auditability, analyst accountability, privacy, cost discipline, and operational control?} This question aligns with IEEE Transactions on Engineering Management because it concerns technology adoption, engineering governance, organizational readiness, and risk-controlled implementation of an emerging capability. The study does not position KQL syntax, a detection rule, or a Microsoft-specific product alternative as the primary contribution. Instead, it studies the management architecture required to convert AI assistance from informal experimentation into an accountable socio-technical capability.

Design science research (DSR) provides the methodological foundation because the study creates and evaluates an artifact intended to improve an organizational capability \cite{hevner2004,peffers2007}. The artifact is a governed AI query broker instantiated in KQL-based threat hunting. It allows AI systems to plan, explain, and propose operational analyses, but it requires executable queries and coding-agent changes to pass through schema grounding, template constraints, policy validation, source-specific adapters, CI/CD gates, human approval where needed, and audit logging.

The article makes four contributions. First, it reframes AI-assisted security operations as an engineering management problem involving governance, risk, cost, accountability, adoption readiness, and evidence maturity. Second, it contributes a DSR artifact that integrates schema-grounded AI planning, brokered read-only execution, agentic tool boundaries, and engineering review board (ERB) promotion gates. Third, it provides design propositions, requirements traceability, operating roles, maturity stages, and quality gates that extend the body of knowledge on governed AI adoption in technical functions. Fourth, it proposes an evaluation protocol that separates package-level engineering validation from stronger claims requiring historical replay, purple-team exercises, shadow-mode SOC use, and limited production deployment. The message for the IEEE TEM community is direct: the value and impact of AI in high-risk engineering work depend less on model fluency than on accountable operating models, evidence gates, and managerial control.

\section{Study Positioning and Research Objectives}
The study is positioned as an engineering management contribution rather than as a purely technical cybersecurity implementation. The empirical context is Microsoft-oriented threat hunting, but the management problem is broader: organizations increasingly adopt AI tools inside technical functions that already depend on privileged access, specialist expertise, software repositories, cloud resources, and audit-sensitive evidence. In such contexts, AI adoption cannot be assessed only by whether a model produces useful text. It must also be assessed by whether the organization can control execution rights, preserve accountability, maintain cost discipline, and generate defensible evidence of safety and value.

KQL and Microsoft Azure security capabilities are used as the study setting because they expose the organizational conditions that make AI governance observable: multiple data owners, source-specific schemas, access permissions, API execution paths, retention limits, query-cost implications, and audit expectations. The technical setting is therefore not incidental, but it is also not the main scholarly claim. It provides a concrete environment in which managers must govern AI-generated work products, while the transferable contribution is the governance and adoption model. The study does not claim to introduce a new query language technique, detection algorithm, or Microsoft product alternative. It uses KQL to instantiate and evaluate a broader management framework for governing AI-assisted operational work in high-risk digital infrastructure.

The artifact developed in this study addresses a class of problems in which AI-generated technical work products can create operational risk if they bypass governance. In the KQL setting, those work products include generated queries, threat-hunting plans, source selections, detection-engineering changes, and analyst-facing interpretations. Similar patterns occur in other engineering management contexts, such as AI-assisted software engineering, infrastructure automation, cloud cost optimization, and safety-critical analytics. The body-of-knowledge contribution is therefore a set of design propositions and managerial control mechanisms for governing AI-assisted work in high-risk technical functions.

Table~\ref{tab:positioning} summarizes the study's positioning. The key point is that the artifact is not evaluated as a replacement for SOC analysts or as a universal detection algorithm. It is evaluated as a governance architecture that helps managers control AI-enabled work across a technical operating model.

\begin{table}[!t]
\caption{Study Positioning in Engineering Management}
\label{tab:positioning}
\centering
\footnotesize
\begin{tabularx}{\columnwidth}{L{0.33\columnwidth}Y}
\toprule
\textbf{Dimension} & \textbf{Position in This Study} \\
\midrule
Primary problem & Managing AI adoption in a privileged, high-risk engineering function. \\
Artifact type & Socio-technical governance artifact with technical, process, role, and evidence components. \\
Unit of contribution & Design propositions, maturity stages, role model, quality gates, and evaluation protocol. \\
Empirical context & Microsoft-centric SOC operations using KQL as a bounded read-only query instantiation. \\
Transferable contribution & Governance, adoption, evidence, and accountability mechanisms for AI-assisted work in high-risk technical functions. \\
Managerial audience & SOC leaders, detection-engineering managers, security architects, AI governance teams, and CISOs. \\
Claim boundary & Design plausibility and controlled evaluation readiness, not completed multi-enterprise production effectiveness. \\
\bottomrule
\end{tabularx}
\end{table}

The research objectives are fourfold. First, the study identifies the managerial controls needed to govern AI-generated security queries. Second, it designs a brokered architecture that separates AI planning from operational execution. Third, it specifies a role-centered operating model for sustaining the artifact. Fourth, it defines evaluation stages that allow managers to decide whether the artifact should remain a proof of concept, progress to shadow mode, or move into controlled production.

\section{Theoretical Background}
\subsection{Engineering Management of AI Adoption}
Engineering management research examines how organizations plan, implement, control, and derive value from technical systems. AI adoption in SOCs fits this domain because leaders must balance productivity gains against risk, accountability, human expertise, process maturity, and technical debt. The management problem is intensified by the dual nature of AI: it can augment human judgment, but it can also automate action without sufficient transparency or control. Raisch and Krakowski frame this tension as an automation-augmentation paradox in AI-enabled organizations \cite{raisch2021}. Berente et al. similarly argue that managing AI requires attention to autonomy, learning, inscrutability, and embeddedness in organizational processes \cite{berente2021}; organizational AI and human-AI work research highlights the need to connect AI capability with decision processes, expertise boundaries, human work, and value realization \cite{jarrahi2018,faraj2018}. These perspectives support a design in which AI is not treated as an independent actor but as a governed component embedded in a controlled work system.

Digital transformation and digital innovation research also show that technology value depends on complementary organizational capabilities, governance, innovation routines, and process change rather than tool installation alone \cite{vial2019,bharadwaj2013,nambisan2017}. Dynamic capabilities theory suggests that organizations must sense opportunities, seize them through investments and structures, and transform routines to sustain value \cite{teece2007}. In this study, the query broker becomes a transformation mechanism: it institutionalizes AI-enabled threat hunting through roles, templates, policy gates, evidence packages, and evaluation routines.

\subsection{AI Governance and Human-AI Collaboration}
Responsible AI governance requires organizations to map AI use contexts, measure risks, manage controls, and govern accountability. NIST's AI Risk Management Framework emphasizes governance, mapping, measurement, and risk management as mutually reinforcing functions \cite{nist_ai_rmf}. In high-risk security operations, these ideas must be operationalized as concrete controls: read-only access, least privilege, prompt and context management, audit logging, approval gates, and evidence separation.

Human-AI collaboration research emphasizes that effective AI systems need calibrated transparency, human control, and role clarity. Guidelines for human-AI interaction argue that systems should make clear what they can do, show uncertainty, support efficient invocation and dismissal, and enable human correction \cite{amershi2019}. Agent transparency research similarly stresses that users need visibility into an agent's goals, reasoning, proposed actions, and uncertainty \cite{voessing2022}. These principles are critical in SOCs because analysts must preserve evidentiary judgment. The AI system should explain why it selected a table, template, entity, or pivot; it should not convert its interpretation into a verified incident finding.

\subsection{AI-Assisted SOC and Detection Engineering}
Recent studies show that large language models (LLMs), retrieval-augmented generation (RAG), and tool-using agents can support SOC triage, anomaly explanation, query generation, and incident response. SERC uses RAG to support Wazuh security-event response \cite{ismail2025serc}. Research on generative AI in cybersecurity identifies threat-hunting query generation as a practical use case \cite{sai2024generative}, while adjacent work on human-AI collaboration shows that AI systems must be designed for effective use, correction, and organizational integration rather than simple tool availability \cite{burtonjones2013,jarrahi2018}. These studies demonstrate technical promise, but they do not fully explain how engineering managers should govern AI-generated security queries, repository modifications, threat-intelligence use, and live telemetry access.

Table~\ref{tab:literature_matrix} summarizes the literature gap. The table shows that prior work provides either general AI management theory, DSR rigor guidance, human-AI design guidance, or SOC-oriented AI prototypes. The present study combines these streams into a governed implementation framework for a specific high-risk engineering function.

\begin{table}[!t]
\caption{Relevant Literature Streams}
\label{tab:literature_matrix}
\centering
\footnotesize
\begin{tabularx}{\textwidth}{L{0.18\textwidth}L{0.25\textwidth}Y Y}
\toprule
\textbf{Literature Stream} & \textbf{Representative Insight} & \textbf{Limitation for This Problem} & \textbf{Contribution of This Study} \\
\midrule
AI management and digital transformation & AI adoption requires organizational capabilities, human augmentation, and management of automation risk \cite{raisch2021,berente2021,vial2019}. & Provides high-level governance insight but not a concrete artifact for privileged security-query execution. & Converts AI management principles into broker controls, maturity stages, and SOC operating roles. \\
Human-AI collaboration & Users need calibrated transparency, uncertainty, correction paths, and visibility into agent actions \cite{amershi2019,voessing2022,miller2019}. & Often focuses on interface guidance rather than organizational promotion gates and execution control. & Embeds transparency in hunt-plan previews, evidence separation, approval decisions, and trace logs. \\
Design science research & DSR provides artifact construction, evaluation, transparency, validity, and reliability guidance \cite{hevner2004,peffers2007,venable2016,hevner2024}. & Requires domain-specific operationalization for AI-enabled SOC systems. & Applies DSR rigor to claim boundaries, evidence maturity, ablations, and reproducibility packages. \\
SOC and AI-security prototypes & RAG copilots and generative AI security applications can support analysts \cite{ismail2025serc,sai2024generative}. & Technical promise is not sufficient for production governance, cost control, and accountability. & Adds policy-enforced execution, template ownership, threat-intelligence provenance, and ERB gates. \\
Agentic and coding-agent systems & Agentic systems can decompose tasks and modify artifacts, but agency creates safety concerns \cite{baird2021,kellogg2020}. & General agentic theory does not specify controls for detection-engineering repositories and live telemetry. & Defines sandboxed coding-agent workflows, tool manifests, CI/CD gates, and prohibited-action tests. \\
\bottomrule
\end{tabularx}
\end{table}

Human-centered SOC research identifies mismatches between tools, managerial expectations, and analyst needs \cite{kokulu2019socs}. Broader organizational AI research similarly shows that algorithmic systems reshape expertise, work boundaries, and decision processes, which supports a human-governed design in which AI assists analysts without becoming the accountable security actor \cite{faraj2018,jarrahi2018}.

\subsection{Design Science and Design Knowledge}
DSR is appropriate because the study produces a purposeful artifact that addresses a relevant organizational problem \cite{hevner2004}. Peffers et al. provide a process model of problem identification, objectives, design and development, demonstration, evaluation, and communication \cite{peffers2007}. Contemporary DSR research adds important rigor criteria. Gregor and Hevner emphasize that DSR work should communicate clear contributions for both research and practice \cite{gregor2013}. Baskerville et al. argue that DSR contributions must balance artifact novelty with design theory contribution \cite{baskerville2018}. Venable et al. provide FEDS as a framework for staging artificial and naturalistic evaluation \cite{venable2016}. Hevner et al. emphasize transparency across process, problem space, solution space, build, evaluation, and contribution \cite{hevner2024}. Recent work on reliability and validity in DSR further stresses the need to align evidence with claim types \cite{storey2025,larsen2025}. This study uses these concepts to avoid unsupported claims about production SOC performance.

\subsection{Research Gap}
The related literature establishes that AI can support security-event interpretation, query generation, and coding assistance. It also establishes that AI adoption requires governance, transparency, and organizational capabilities. However, there is limited design knowledge explaining how managers should introduce AI-generated security queries into privileged telemetry environments without granting models or agents uncontrolled authority. The gap is not only technical. It concerns how organizations structure roles, evidence gates, policy controls, evaluation stages, and repository workflows so that AI-enabled security operations remain auditable, cost-aware, and accountable. This study addresses that gap through a DSR artifact and a management-oriented evaluation framework.

\subsection{Conceptual Model}
The conceptual model links five constructs: AI capability, governance control, operational integration, evidence maturity, and managerial value. AI capability includes natural-language interpretation, schema-grounded retrieval, query generation, explanation, and coding-agent assistance. Governance control includes policy validation, least privilege, approval gates, source allowlists, audit logging, and repository controls. Operational integration includes role assignment, workflow fit, template ownership, and CI/CD. Evidence maturity describes the quality of evidence available to justify deployment decisions. Managerial value includes safer experimentation, faster triage, more consistent detection engineering, improved auditability, and better cost visibility.

The model assumes that AI capability alone is insufficient. Without governance controls, stronger AI capability can increase risk by making it easier to generate broad, persuasive, or unsafe queries. Governance controls moderate the relationship between AI capability and managerial value. Operational integration mediates the relationship because AI-enabled artifacts create value only when they fit analyst work, repository practice, and approval processes. Evidence maturity moderates adoption decisions because the same artifact should support different conclusions at different evidence levels.

\begin{table}[!t]
\caption{Conceptual Constructs and Observable Indicators}
\label{tab:constructs}
\centering
\footnotesize
\begin{tabularx}{\columnwidth}{L{0.28\columnwidth}Y}
\toprule
\textbf{Construct} & \textbf{Observable Indicators} \\
\midrule
AI capability & Valid hunt plans, grounded schema retrieval, executable KQL candidates, useful explanations, and coding-agent diffs. \\
Governance control & Policy decisions, blocked unsafe queries, approval workflow records, source allowlists, and audit completeness. \\
Operational integration & Assigned roles, template ownership, CI/CD checks, training records, and analyst workflow fit. \\
Evidence maturity & Package tests, replay outcomes, purple-team results, shadow-mode pilot data, and production monitoring. \\
Managerial value & Reduced unsafe query variance, faster evidence discovery, better documentation, cost visibility, and defensible adoption decisions. \\
\bottomrule
\end{tabularx}
\end{table}

\subsection{Boundary Conditions}
The conceptual model has explicit boundary conditions. It applies most directly to organizations with a mature security telemetry platform, defined SOC roles, read-only investigation workflows, detection-engineering ownership, and security review processes. It is less applicable to organizations that lack basic telemetry onboarding, identity governance, or incident-response procedures. The artifact also assumes that response actions are outside the initial scope. Device isolation, account disablement, email purge, or detection-rule deployment should require separate governance because those actions create higher operational impact than read-only hunting.

\section{Research Design}
\subsection{Design Science Process}
The study follows DSRM and treats the query broker as a socio-technical artifact. Table~\ref{tab:dsr} maps the study to DSR activities. The artifact is evaluated conceptually and through engineering validation criteria. Stronger organizational-performance claims are reserved for future naturalistic evaluation.

\begin{table}[!t]
\caption{Design Science Research Mapping}
\label{tab:dsr}
\centering
\footnotesize
\begin{tabularx}{\columnwidth}{L{0.31\columnwidth}Y}
\toprule
\textbf{DSR Activity} & \textbf{Application in This Study} \\
\midrule
Problem identification & AI can generate useful KQL, but unmanaged execution can create privilege, privacy, cost, and audit risk. \\
Objectives & Support AI-assisted threat hunting while enforcing least privilege, schema validity, bounded execution, human accountability, and auditability. \\
Design and development & Construct a broker architecture with RAG grounding, approved templates, KQL validation, source adapters, agent traces, CI/CD gates, and ERB promotion criteria. \\
Demonstration & Apply the design to Microsoft Defender XDR, Sentinel, Sentinel data lake, and ADX-oriented hunting scenarios. \\
Evaluation & Specify offline validation, historical replay, purple-team simulation, shadow-mode SOC pilot, and limited production evidence. \\
Communication & Provide a management-oriented artifact, propositions, requirements traceability, maturity model, and implementation guidance. \\
\bottomrule
\end{tabularx}
\end{table}

\subsection{Research Questions}
The study addresses three research questions:
\begin{enumerate}
    \item RQ1: What governance mechanisms are needed to manage AI-assisted operational decision support in high-risk SOC environments?
    \item RQ2: How can AI planning be separated from controlled read-only query execution and coding-agent activity?
    \item RQ3: How should managers evaluate such an artifact before progressing from proof of concept to shadow mode or production use?
\end{enumerate}

\subsection{Evidence Boundary}
The study is a design and evaluation-protocol contribution. It does not claim that production SOC detection lift, analyst adoption, or cost reduction has been demonstrated across multiple enterprises. Current evidence can support design plausibility, engineering review readiness, reproducibility, and controlled pilot suitability. Production-effectiveness claims require historical incident replay, purple-team exercises, shadow-mode SOC pilots, and limited production deployment.

This boundary matters because AI demonstrations can create premature confidence. Package-level validation means that the artifact can be inspected, compiled, linted, and tested against synthetic or schema-normalized data. It does not mean that an organization has achieved production detection lift. Naturalistic evidence is required before leaders use the artifact to justify staffing reductions, automated response, or broad operational deployment.

\subsection{Design Propositions}
The study develops six design propositions that connect the artifact to generalizable engineering management knowledge.

\textbf{P1: Brokered execution.} In privileged telemetry environments, separating AI planning from executable access increases managerial control by converting AI output into a policy-evaluated request rather than an autonomous action.

\textbf{P2: Template-first generation.} AI-generated security queries are more governable when they derive from approved parameterized templates with owners, metadata, tests, and version history.

\textbf{P3: Provenance-aware intelligence.} Threat-intelligence use becomes more reliable when indicators preserve source, confidence, validity, and handling constraints before influencing query generation or risk scoring.

\textbf{P4: Evidence-bounded evaluation.} Engineering managers can reduce premature adoption risk by matching each claim to an appropriate evidence level, from offline tests to naturalistic deployment.

\textbf{P5: Sandboxed agentic contribution.} Coding agents can support detection engineering when restricted to nonproduction repositories, pull requests, CI/CD gates, and human review.

\textbf{P6: Role-centered governance.} AI adoption in SOCs is more sustainable when responsibilities are assigned across SOC leadership, detection engineering, platform engineering, threat intelligence, and AI governance rather than centralized in the model provider or analyst alone.

\section{Artifact Design}
\subsection{Architecture Overview}
\begin{figure}[!t]
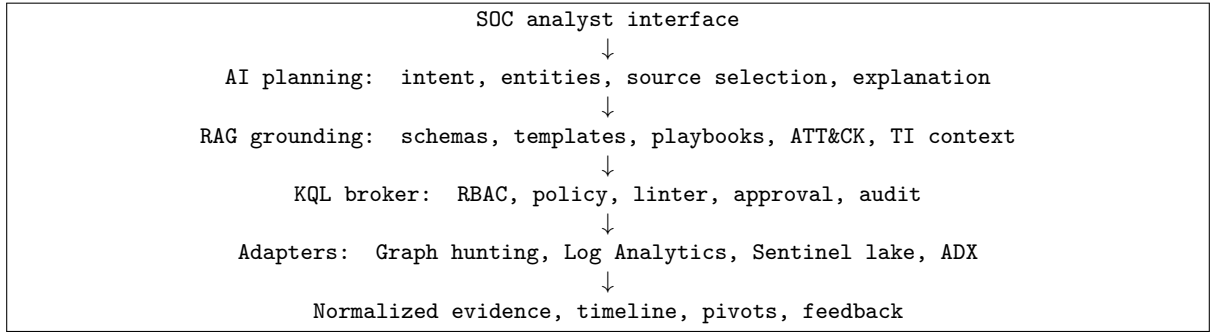

\centering
\footnotesize
\fbox{\begin{minipage}{0.95\columnwidth}
\centering\ttfamily
SOC analyst interface\\
$\downarrow$\\
AI planning: intent, entities, source selection, explanation\\
$\downarrow$\\
RAG grounding: schemas, templates, playbooks, ATT\&CK, TI context\\
$\downarrow$\\
KQL broker: RBAC, policy, linter, approval, audit\\
$\downarrow$\\
Adapters: Graph hunting, Log Analytics, Sentinel lake, ADX\\
$\downarrow$\\
Normalized evidence, timeline, pivots, feedback
\end{minipage}}
\caption{Governed AI query-broker architecture for enterprise threat hunting.}
\label{fig:architecture}
\end{figure}

The artifact is a governed query broker that separates natural-language AI assistance from executable security-data access. The core control flow is: analyst request, AI hunt plan, RAG grounding, approved template selection, KQL validation, policy decision, read-only API execution, result normalization, explanation, and analyst feedback. Fig.~\ref{fig:architecture} summarizes the architecture.

The architecture follows a management principle: models may propose and explain, but the broker decides whether a query can execute. This separation reduces excessive agency, protects privileged data sources, and gives managers an auditable control point.

\subsection{Component Contracts}
The artifact contains six management-relevant components. The AI planner converts analyst intent into a structured hunt plan. The RAG layer retrieves schemas, templates, playbooks, MITRE ATT\&CK mappings, and threat-intelligence context. The template resolver maps the plan to an approved KQL pattern. The broker evaluates the query against identity, policy, source, cost, time, row-limit, and approval rules. Source adapters isolate Microsoft Graph, Log Analytics, Sentinel data lake, and ADX execution. The normalization and feedback layer converts source-specific results into common evidence objects and records analyst disposition.

\begin{table}[!t]
\caption{Component Contracts and Managerial Controls}
\label{tab:components}
\centering
\footnotesize
\begin{tabularx}{\columnwidth}{L{0.25\columnwidth}Y Y}
\toprule
\textbf{Component} & \textbf{Output} & \textbf{Managerial Control} \\
\midrule
AI planner & Structured hunt plan and rationale. & Plan preview and human-visible assumptions. \\
RAG grounding & Retrieved schemas, templates, and context. & Versioned knowledge base and stale-context monitoring. \\
Template resolver & Parameterized KQL candidate. & Template ownership, review status, and metadata. \\
KQL broker & Approve, modify, reject, or escalate decision. & Policy enforcement, RBAC, cost limits, and audit. \\
API adapters & Source-specific read-only results. & Least-privilege identities and adapter allowlists. \\
Feedback layer & Analyst disposition and evaluation labels. & Learning constraints and improvement backlog. \\
\bottomrule
\end{tabularx}
\end{table}

\subsection{Requirements Traceability}
Table~\ref{tab:req} links requirements to design mechanisms and evidence. This traceability helps readers assess whether the artifact addresses the stated management problem rather than presenting disconnected technical components.

\begin{table}[!t]
\caption{Requirements, Design Mechanisms, and Evaluation Evidence}
\label{tab:req}
\centering
\footnotesize
\begin{tabularx}{\textwidth}{L{0.17\textwidth}L{0.30\textwidth}Y Y}
\toprule
\textbf{Requirement} & \textbf{Management Intent} & \textbf{Design Mechanism} & \textbf{Evaluation Evidence} \\
\midrule
Grounded generation & Prevent plausible but invalid KQL. & RAG over current schemas, approved templates, and source-specific conventions. & Schema-validity rate, invalid-column rate, table-selection accuracy. \\
Controlled execution & Prevent direct AI access to production telemetry. & Brokered API execution, read-only adapters, source allowlists, and managed identities. & Policy-bypass tests, least-privilege review, audit-log completeness. \\
Operational cost control & Avoid unbounded scans and throttling. & Required time filters, row limits, query budgets, and cost monitoring. & Latency, timeout rate, throttling rate, cost per hunt. \\
Human accountability & Keep analysts and managers responsible for disposition. & Hunt-plan preview, approval gates, explanation, uncertainty markers, and feedback capture. & Analyst comprehension, accepted-pivot rate, usefulness ratings. \\
Agentic containment & Allow coding support without repository or telemetry risk. & Sandboxed coding agents, pull requests, CI/CD checks, no secrets, and no live APIs. & Prohibited-tool blocking, reviewer acceptance, linter/test pass rate. \\
Reproducibility & Make the design inspectable without live telemetry. & Synthetic data, public mapping logic, versioned prompts, templates, policies, and manifests. & Package validation, expected-output tests, repeat benchmark consistency. \\
\bottomrule
\end{tabularx}
\end{table}

\subsection{Design Principles}
The artifact is governed by eight principles. First, AI assistance must be brokered before execution. Second, KQL must be grounded in current schemas and approved templates. Third, every production-facing query must be time-bounded and result-bounded, consistent with KQL performance guidance to reduce processed data early \cite{ms_kql_best_practices}. Fourth, threat intelligence must preserve provenance, confidence, TLP handling, and validity windows. Current Sentinel threat-intelligence guidance emphasizes STIX-oriented tables such as \texttt{ThreatIntelIndicators} and \texttt{ThreatIntelObjects} rather than new development around legacy indicator assumptions \cite{ms_sentinel_ti}. Fifth, explanations must separate observed evidence from AI interpretation. Sixth, coding agents must operate in sandboxed repository workflows rather than production operations. Seventh, promotion from proof of concept to production must depend on ERB evidence, not on model fluency. Eighth, managers must treat evaluation as a staged risk-control process rather than a single benchmark.

\begin{table}[!t]
\caption{Design Principles and Managerial Control Interpretation}
\label{tab:principles}
\centering
\footnotesize
\begin{tabularx}{\textwidth}{L{0.19\textwidth}L{0.34\textwidth}Y}
\toprule
\textbf{Principle} & \textbf{Technical Meaning} & \textbf{Managerial Interpretation} \\
\midrule
Brokered execution & Every executable query passes through policy validation and source-specific adapters. & AI assistance is an input to a governed workflow, not an independent operator. \\
Schema grounding & The AI retrieves current table, column, and source metadata before producing KQL. & Managers should fund schema refresh and data-catalog ownership. \\
Template-first generation & Production-facing KQL derives from reviewed parameterized templates. & Detection content becomes an engineering asset with owners and version history. \\
Least privilege & Broker identities and human roles receive only needed read-only access. & AI adoption should not expand standing privileges or bypass RBAC. \\
Time and row bounding & Queries require lookback limits, projections, and row caps. & Cost and performance risk are managed before live execution. \\
Threat-intelligence provenance & Indicator joins retain source, confidence, validity, and handling data. & OSINT should inform hypotheses but not become unverified proof. \\
Agentic containment & Coding agents work through pull requests, tests, and human review. & AI can accelerate engineering work without controlling production repositories. \\
Evidence-bounded claims & Each claim is tied to an evidence level and a decision gate. & Leaders avoid using POC results to justify broad production claims. \\
\bottomrule
\end{tabularx}
\end{table}

\subsection{KQL as a Bounded Technical Instantiation}
Listing~\ref{lst:kql} shows a simplified example of the template-first approach. The listing is included to make the governance mechanism technically inspectable, not to claim novelty in KQL syntax. The example demonstrates an explicit lookback, dynamic parsing of extended fields, projection control, and result bounding. These controls matter even though ordinary KQL queries are read-only because poorly scoped read-only queries can still expose sensitive evidence, increase cost, create misleading investigative pivots, or undermine auditability. In production, values such as allowed tasks, sensitive accounts, and enterprise domains should come from governed watchlists rather than hard-coded prompts. This framing keeps the study's contribution at the engineering management level: how AI-generated operational artifacts are proposed, validated, approved, executed, audited, and improved.

\begin{lstlisting}[language=KQL,caption={Illustration of scheduled-task hunting template with bounded execution.},label={lst:kql}]
let Lookback = 7d;
DeviceEvents
| where Timestamp > ago(Lookback)
| where ActionType == "ScheduledTaskCreated"
| extend Ext = todynamic(AdditionalFields)
| extend TaskName = tostring(Ext.TaskName),
         TaskContent = tostring(Ext.TaskContent)
| where InitiatingProcessAccountName !endswith "$"
| project Timestamp, DeviceName, InitiatingProcessAccountName,
          InitiatingProcessFileName,
          InitiatingProcessCommandLine,
          TaskName, TaskContent, ReportId
| top 100 by Timestamp desc
\end{lstlisting}

\subsection{Traceability From Request to Evidence}
A key management contribution is end-to-end traceability. The artifact should record the initiating request, user identity, retrieved context, selected template, generated KQL, policy decision, approvals, execution metadata, normalized results, explanation, and analyst disposition. This trace creates accountability and supports post-incident review, model-risk monitoring, and continuous improvement. It also allows managers to distinguish three failure classes: failures of AI planning, failures of governance control, and failures of operational fit.

\begin{table}[!t]
\caption{Traceability Chain for Governed AI-Enabled Threat Hunting}
\label{tab:trace}
\centering
\footnotesize
\begin{tabularx}{\textwidth}{L{0.15\textwidth}L{0.25\textwidth}Y Y}
\toprule
\textbf{Trace Stage} & \textbf{Recorded Evidence} & \textbf{Managerial Use} & \textbf{Failure Signal} \\
\midrule
Request intake & User, role, stated objective, time window, source request, and case context. & Confirms legitimate purpose and accountable requester. & Unclear objective, excessive scope, or unsupported source. \\
Grounding & Retrieved schemas, templates, playbooks, and threat-intelligence context identifiers. & Shows whether AI output was grounded in approved knowledge. & Stale schema, weak retrieval, or missing template. \\
Generation & Candidate KQL, parameters, assumptions, and explanation. & Supports review of AI reasoning and technical fit. & Hallucinated table, missing field, or unsupported assumption. \\
Broker decision & Policy outcome, reason, approval state, limits, and modified query if applicable. & Demonstrates control effectiveness and audit defensibility. & Policy bypass, missing approval, or unexplained modification. \\
Execution & Adapter, source, start time, latency, row count, and error state. & Supports cost, reliability, and source-use monitoring. & Timeout, throttling, excessive row return, or repeated errors. \\
Disposition & Analyst feedback, accepted pivots, rejected recommendations, and case notes. & Supports learning, training, and content improvement. & Overreliance, low usefulness, or repeated false-positive pattern. \\
\bottomrule
\end{tabularx}
\end{table}

\subsection{Microsoft Security Data Context}
The artifact is grounded in a Microsoft-centric telemetry environment but abstracts platform differences through adapters. Defender XDR advanced hunting is useful for endpoint, identity, email, and cloud-app investigation; Sentinel and Log Analytics support SIEM-style workspace queries; Sentinel data lake supports long-retention analytics and historical investigation; and Azure Data Explorer supports custom telemetry and scalable analytics. This setting is suitable for an engineering management study because each source has different schemas, retention patterns, permissions, cost behavior, and operational ownership. A broker gives managers a single control plane for applying source policy while preserving source-specific execution. The Microsoft/KQL context is therefore a realistic and bounded instantiation of a broader problem: governing AI assistance in high-risk data-intensive engineering work.

\begin{table}[!t]
\caption{Security Data Sources and Management Implications}
\label{tab:sources}
\centering
\footnotesize
\begin{tabularx}{\columnwidth}{L{0.30\columnwidth}Y}
\toprule
\textbf{Source} & \textbf{Management Implication} \\
\midrule
Defender XDR advanced hunting & Requires attention to Graph permissions, row limits, schema changes, and endpoint/identity/email table ownership. \\
Sentinel and Log Analytics & Requires workspace RBAC, cost monitoring, retention policy, custom table governance, and rule ownership. \\
Sentinel data lake & Requires stronger data-retention, historical search, and long-running query governance. \\
Azure Data Explorer & Requires cluster cost governance, database permissions, and custom schema catalog management. \\
Threat intelligence tables & Require source confidence, validity windows, TLP handling, deduplication, and analyst corroboration. \\
\bottomrule
\end{tabularx}
\end{table}

\section{Operating Model and Implementation Pathway}
\subsection{Operating Roles}
The artifact requires a multidisciplinary operating model. SOC analysts use the interface to submit hunts, review plans, interpret results, and provide feedback. Detection engineers own KQL templates, tests, and promotion decisions. Security architects own identities, network boundaries, adapter access, and platform integration. Threat-intelligence analysts own source quality, confidence, and handling rules. Data and platform engineers own schema mappings, synthetic datasets, pipelines, and observability. AI governance leads own model-risk controls, prompt policies, traceability, and responsible-use review. The executive sponsor owns risk appetite, funding, and production approval criteria.

\begin{table}[!t]
\caption{Operating Roles and Accountability}
\label{tab:roles}
\centering
\footnotesize
\begin{tabularx}{\columnwidth}{L{0.31\columnwidth}Y}
\toprule
\textbf{Role} & \textbf{Primary Accountability} \\
\midrule
CISO or executive sponsor & Risk appetite, funding, adoption roadmap, and production approval. \\
SOC manager & Workflow fit, analyst training, escalation boundaries, and pilot success criteria. \\
Detection engineering lead & Template ownership, benchmark cases, linter rules, and production promotion. \\
Security architect & RBAC, managed identity, network boundary, adapter scope, and key management. \\
Threat-intelligence lead & Source provenance, confidence, validity, TLP handling, and enrichment policy. \\
AI governance lead & Model-risk controls, prompt policy, agent tool manifest, trace review, and responsible use. \\
Data/platform engineer & Schema mapping, synthetic data, IaC, CI/CD, monitoring, and reliability dashboards. \\
\bottomrule
\end{tabularx}
\end{table}

\subsection{Maturity Model}
The design should be implemented through staged maturity rather than immediate production deployment. Stage 0 is offline review with synthetic data and mock adapters. Stage 1 is a controlled proof of concept using approved templates, package validation, and no live telemetry. Stage 2 is a shadow-mode pilot using limited read-only telemetry and analyst review. Stage 3 is controlled production use with trained analysts, audited execution, and budget controls. Stage 4 extends the model to coding-agent support for detection engineering, still through pull requests and human review. Table~\ref{tab:maturity} defines the evidence expected at each stage.

\begin{table}[!t]
\caption{Maturity Stages, Evidence Gates, and Managerial Decisions}
\label{tab:maturity}
\centering
\footnotesize
\begin{tabularx}{\textwidth}{L{0.11\textwidth}L{0.25\textwidth}Y Y}
\toprule
\textbf{Stage} & \textbf{Capability Boundary} & \textbf{Evidence Required} & \textbf{Managerial Decision} \\
\midrule
0. Offline review & Synthetic data, mock adapters, documented templates, no live telemetry. & Architecture, data-use boundary, source manifest, KQL templates, policy file, and package checksums. & Approve or reject POC investment. \\
1. Controlled POC & Broker validation and approved templates in a nonproduction environment. & KQL linter results, schema checks, unsafe-query blocking, prompt-injection tests, and reproducibility package. & Approve or revise shadow-mode pilot. \\
2. Shadow-mode pilot & Limited read-only telemetry with analyst oversight and no automated response. & Pilot training, audit logs, analyst usefulness, timing data, false-positive notes, and cost observations. & Approve controlled production release or restrict use cases. \\
3. Controlled production & Approved analysts execute brokered read-only hunts under budgets and monitoring. & Reliability, cost per hunt, latency, policy-bypass attempts, audit completeness, and incident contribution evidence. & Expand, continue, or roll back production use. \\
4. Agent-assisted engineering & Coding agents propose template, policy, test, and documentation changes through pull requests. & CI/CD pass rate, reviewer acceptance, prohibited-tool blocking, trace completeness, and rollback readiness. & Permit broader detection-engineering support under repository governance. \\
\bottomrule
\end{tabularx}
\end{table}

\subsection{Engineering Review Board Gates}
The ERB gate structure converts abstract governance into reviewable evidence. Gate 0 approves offline POC review when the architecture, data boundary, threat model inputs, and reproducibility package exist. Gate 1 approves shadow-mode pilot when read-only identity, source allowlists, broker policy, analyst training, monitoring, and pilot success criteria are complete. Gate 2 approves controlled production when pilot outcomes, cost reports, latency reports, audit completeness, RACI, change management, and residual-risk acceptance are available. Stop conditions include excessive privileges, missing audit logs, failed secret scanning, live production write capability, unstable cost, unresolved privacy concerns, or inability to roll back.

\subsection{DevSecOps Quality Gates}
The artifact should be managed as secure software. KQL templates, prompt files, RAG schemas, policy files, adapter code, evaluation scripts, and infrastructure-as-code should move through protected branches, code-owner review, secret scanning, dependency scanning, IaC scanning, KQL linting, prompt-injection tests, and release manifests. These controls are consistent with secure software development guidance \cite{nist_ssdf} and application-security maturity principles \cite{owasp_samm}. Coding agents may create candidate changes, but they may not merge, deploy, access secrets, or call production APIs. This operating rule converts agentic support into a controlled engineering contribution.

\begin{table}[!t]
\caption{Repository and CI/CD Controls}
\label{tab:cicd}
\centering
\footnotesize
\begin{tabularx}{\columnwidth}{L{0.28\columnwidth}Y}
\toprule
\textbf{Gate} & \textbf{Block Condition} \\
\midrule
Static validation & Invalid YAML, invalid KQL metadata, missing owner, unresolved citation, or missing schema version. \\
Security scanning & Exposed secret, over-permissive IaC, critical dependency risk, or unapproved network path. \\
Policy testing & Expanded table access, missing time filter, missing row limit, or disabled approval rule. \\
Benchmark testing & Expected-output regression, increased hallucination rate, or failed unsafe-prompt blocking. \\
Release packaging & Missing manifest, missing checksum, incomplete changelog, or absent rollback instructions. \\
\bottomrule
\end{tabularx}
\end{table}

\subsection{Governance Walkthrough}
A governed workflow clarifies how the artifact changes day-to-day SOC operations. Consider a suspicious scheduled-task alert. In an ungoverned AI workflow, an analyst may paste the alert into an LLM and receive a free-form KQL query that may or may not match tenant schemas or source permissions. In the proposed workflow, the analyst opens a brokered hunt session, selects a case and approved time window, and describes the investigative hypothesis. The AI planner generates a structured plan, but the plan is not executable. The RAG layer retrieves the approved scheduled-task template, relevant schema metadata, and applicable playbook context. The template resolver binds permitted parameters. The broker then validates the query, checks source permissions, confirms the row cap, applies sensitive-field rules, and records the decision. Only then does the adapter execute a read-only query.

The managerial benefit of this workflow is not only faster query generation. It is reduced variance in how analysts use AI, a complete audit trail, and a defensible explanation of why a query was allowed. If the query is blocked, the policy decision becomes a learning artifact: either the analyst request exceeded policy, the template library lacks a needed approved hunt, or the source boundary needs governance review. Thus, blocked actions create management information rather than hidden friction.

\subsection{Risk Register}
The implementation should maintain a risk register that links technical failure modes to management controls. Table~\ref{tab:risk_register} summarizes representative risks. The risk register is important for TEM because it converts a cybersecurity prototype into a managed engineering capability with ownership, mitigations, and decision triggers.

\begin{table}[!t]
\caption{Risk Register for Governed AI-Enabled Threat Hunting}
\label{tab:risk_register}
\centering
\footnotesize
\begin{tabularx}{\textwidth}{L{0.17\textwidth}L{0.27\textwidth}Y Y}
\toprule
\textbf{Risk} & \textbf{Failure Mode} & \textbf{Control Mechanism} & \textbf{Management Trigger} \\
\midrule
Schema hallucination & AI references unavailable tables or fields. & RAG grounding, schema snapshots, template-first generation, and linter tests. & Pause model or prompt promotion when invalid-column rate exceeds threshold. \\
Excessive data scan & Query omits time bounds or returns too many rows. & Required lookback, row caps, projection rules, and query budgets. & Escalate if timeout, throttling, or cost exceeds approved pilot limits. \\
Sensitive data exposure & Query projects raw content or restricted identifiers. & Sensitive-field policy, role-specific output minimization, and approval gates. & Stop production expansion if sensitive-field suppression fails. \\
Weak intelligence use & Low-confidence or expired indicators drive severity. & Provenance, confidence, validity, and corroboration requirements. & Review feed quality if TI-driven false positives exceed target. \\
Prompt injection & Telemetry content instructs model behavior. & Treat logs as evidence, not instructions; isolate context; test adversarial prompts. & Disable affected prompt path until containment tests pass. \\
Agentic overreach & Coding agent attempts to merge, deploy, or access secrets. & Sandboxed workspace, tool manifest, no production APIs, and CI/CD gates. & Revoke tool access after prohibited-action attempt. \\
Analyst overreliance & Users accept AI interpretation without evidence review. & Evidence-interpretation separation, uncertainty labels, and training. & Retrain or restrict use if trust-calibration checks fail. \\
Unclear accountability & Teams disagree on template, policy, or incident ownership. & RACI model, ERB gates, and change-management records. & Block promotion when owners or escalation paths are missing. \\
\bottomrule
\end{tabularx}
\end{table}

\subsection{Adoption Roadmap}
A realistic adoption roadmap should sequence technology, process, and people. The first step is to inventory telemetry sources, schema owners, and current KQL repositories. The second step is to convert ad hoc queries into approved templates with metadata, expected outputs, and tests. The third step is to implement a mock broker with synthetic data so that policy logic can be inspected before live access. The fourth step is to run shadow-mode pilots with trained analysts and limited read-only telemetry. The fifth step is to expand use cases only after the organization has stable cost monitoring, audit retention, and incident-management integration.

The roadmap also requires workforce development. Analysts need training on how to inspect generated KQL, interpret uncertainty, reject unsupported AI claims, and provide feedback. Detection engineers need training on template ownership and regression testing. Platform engineers need training on source adapters, managed identities, and cost monitoring. AI governance teams need training on prompt-risk assessment, agent traces, and model-version evaluation. The artifact therefore changes management practice by making AI-enabled security operations a cross-functional capability rather than a point solution.

\subsection{Managerial Acceptance Criteria}
Before moving from proof of concept to production, managers should require a minimum acceptance package. It should include: an architecture diagram, data-flow diagram, managed-identity design, table allowlist, broker policy, template library, evidence of KQL validation, prompt-injection test results, audit-log sample, cost estimate, rollback procedure, RACI matrix, analyst training plan, and residual-risk statement. These artifacts allow decision makers to compare benefits against risks and to document why the deployment is acceptable.

\begin{table}[!t]
\caption{Minimum Acceptance Package for Production Consideration}
\label{tab:acceptance}
\centering
\footnotesize
\begin{tabularx}{\columnwidth}{L{0.33\columnwidth}Y}
\toprule
\textbf{Evidence Item} & \textbf{Decision Purpose} \\
\midrule
Architecture and data-flow diagrams & Confirm boundary, sources, identities, and data movement. \\
Broker policy and table allowlist & Confirm least privilege and execution constraints. \\
Template library and test results & Confirm governed detection-engineering content. \\
Prompt and agent safety tests & Confirm resistance to unsafe instructions and excessive agency. \\
Audit-log sample & Confirm traceability from request to execution and disposition. \\
Cost and latency report & Confirm operational feasibility under expected use. \\
RACI and training plan & Confirm accountability and user readiness. \\
Rollback and residual-risk statement & Confirm that deployment can be reversed and approved risk is explicit. \\
\bottomrule
\end{tabularx}
\end{table}

\subsection{Comparative Managerial Value}
The brokered design creates value relative to three common alternatives. A manual-only workflow preserves human control but can be slow, inconsistent, and hard to scale across junior and senior analysts. A static query-library workflow improves consistency but may not adapt to changing hypotheses or schemas. An unbrokered AI workflow is flexible but can introduce uncontrolled execution, hidden assumptions, and audit gaps. The proposed brokered workflow attempts to combine the flexibility of AI with the accountability of reviewed templates and policy enforcement. Its value should therefore be judged not only by time savings but by the ratio of useful assistance to governance burden.

\subsection{Implementation Readiness Checklist}
An implementation team should not treat the artifact as ready because the prototype compiles or because generated queries appear plausible. Readiness should be judged through a checklist that covers management, technical, and human factors. The checklist in Table~\ref{tab:readiness_checklist} supports decision consistency across review boards and pilot teams. It also gives future researchers a baseline for comparing readiness assessments across organizations.

\begin{table}[!t]
\caption{Implementation Readiness Checklist}
\label{tab:readiness_checklist}
\centering
\footnotesize
\begin{tabularx}{\columnwidth}{L{0.32\columnwidth}Y}
\toprule
\textbf{Checklist Area} & \textbf{Required Condition} \\
\midrule
Problem fit & The use case requires cross-source hunting and cannot be solved adequately through static dashboards alone. \\
Data readiness & Required telemetry sources, retention periods, and schema owners are known. \\
Control readiness & Broker policies, source allowlists, and approval rules are documented and tested. \\
People readiness & Analysts, detection engineers, platform owners, and AI governance reviewers understand their roles. \\
Evidence readiness & Baseline results, benchmark cases, and pilot metrics are available before expansion. \\
Operational readiness & Monitoring, cost controls, incident escalation, and rollback procedures are defined. \\
\bottomrule
\end{tabularx}
\end{table}

\section{Evaluation Framework}
\subsection{Evaluation Logic}
The evaluation framework is designed to assess utility, safety, reliability, and projectability. Because the artifact is socio-technical, evaluation must address both component behavior and organizational use. The study uses staged evaluation: offline benchmark, historical incident replay, controlled purple-team simulation, shadow-mode SOC pilot, and limited production deployment. This sequence follows the logic of artificial-to-naturalistic evaluation in DSR \cite{venable2016}.

\subsection{Baselines}
Four baselines are needed. The first is manual analyst-written KQL, representing expert human performance. The second is a static approved query library, representing governance without AI flexibility. The third is unbrokered AI-generated KQL, representing flexibility with weak control. The fourth is the full brokered architecture, representing governed AI assistance. Matched tasks should use the same telemetry windows, scenario briefs, source availability, and scoring rubric. If analyst expertise varies, the study should use within-subject comparisons or block analysts by experience.

\subsection{Metrics}
Evaluation should align metrics with task type. Query generation should be evaluated through syntax validity, schema validity, execution success, time-filter compliance, and projection compliance. Retrieval should be evaluated with precision@k, recall@k, mean reciprocal rank, and normalized discounted cumulative gain \cite{jarvelin2002}. Detection scenarios should use precision, recall, F1, false-positive burden, PR-AUC, and calibration when scores are available \cite{fawcett2006,davis2006,guo2017}. Operational outcomes should use time-to-first-useful-evidence, triage time, cost per hunt, latency, throttling, and analyst usefulness. Agentic safety should use prohibited-tool blocking, sandbox containment, trace completeness, and reviewer acceptance.

\subsection{Construct Operationalization}
The evaluation should operationalize constructs at three levels: technical artifact behavior, human work outcomes, and management controls. This separation prevents the study from using a single technical score as a proxy for organizational value. For example, a model can produce syntactically valid KQL while still creating poor managerial outcomes if it increases cost, obscures accountability, or encourages analyst overreliance. Conversely, a conservative broker can improve governance but fail to deliver value if it blocks too many legitimate analyst tasks. Table~\ref{tab:operationalization} summarizes construct operationalization.

\begin{table}[!t]
\caption{Construct Operationalization for Empirical Evaluation}
\label{tab:operationalization}
\centering
\footnotesize
\begin{tabularx}{\textwidth}{L{0.15\textwidth}L{0.24\textwidth}Y Y}
\toprule
\textbf{Construct} & \textbf{Definition} & \textbf{Measures} & \textbf{Data Source} \\
\midrule
Query validity & Degree to which generated KQL is executable and schema-aligned. & Syntax pass, schema pass, required time filter, projection compliance, and execution success. & Linter logs, mock-adapter runs, and source-adapter results. \\
Governance effectiveness & Degree to which controls prevent unsafe or unauthorized use. & Blocked unsafe prompts, approval enforcement, sensitive-field suppression, and audit completeness. & Broker decision logs and seeded adversarial tests. \\
Analyst usefulness & Degree to which outputs support analyst work without excessive burden. & Usefulness rating, accepted-pivot rate, time to useful evidence, and rejected-recommendation reasons. & Shadow-mode tasks, surveys, and analyst annotations. \\
Cost and reliability & Degree to which the artifact operates within operational limits. & P95 latency, timeout rate, throttling, query volume, token consumption, and cost per hunt. & Telemetry dashboards and cloud cost reports. \\
Agentic safety & Degree to which coding agents remain inside authorized tool and repository boundaries. & Prohibited-tool blocking, CI/CD pass rate, reviewer acceptance, and trace completeness. & Repository logs, pull requests, and agent tool traces. \\
Management readiness & Degree to which the organization can operate and govern the artifact. & RACI completeness, training completion, ERB evidence, rollback readiness, and risk acceptance. & Governance artifacts and ERB records. \\
\bottomrule
\end{tabularx}
\end{table}

\subsection{Sampling and Study Design for Future Empirical Work}
A rigorous empirical study should sample both tasks and users. Task sampling should include common hunts, rare high-risk hunts, ambiguous prompts, benign administrative behavior, threat-intelligence enrichment, and adversarial prompt-injection attempts. User sampling should include junior analysts, senior hunters, detection engineers, SOC managers, and platform engineers because each group evaluates different aspects of value. Junior analysts may benefit from scaffolding, senior hunters may focus on flexibility and evidence quality, and managers may focus on auditability, cost, and policy compliance.

A within-subject design is appropriate for analyst workflow studies because the same analyst can perform matched tasks using manual KQL, static templates, unbrokered AI, and the brokered artifact. Order effects should be counterbalanced. Scenarios should be randomized where feasible. Evaluators should record task completion time, query count, analyst confidence, evidence quality, and post-task comments. For qualitative analysis, comments should be coded for perceived usefulness, trust calibration, workflow friction, transparency, and accountability concerns.

\begin{table}[!t]
\caption{Evaluation Stages, Metrics, and Decision Rules}
\label{tab:eval}
\centering
\footnotesize
\begin{tabularx}{\textwidth}{L{0.16\textwidth}L{0.24\textwidth}Y Y}
\toprule
\textbf{Stage} & \textbf{Primary Metrics} & \textbf{Target Decision Evidence} & \textbf{Claim Supported} \\
\midrule
Offline benchmark & KQL validity, schema validity, unsafe-query blocking, retrieval precision, prompt-injection resistance. & At least 95\% schema-valid KQL, 100\% required time filters, and 100\% blocking of seeded unsafe requests. & Artifact readiness for controlled pilot, not production SOC effectiveness. \\
Historical replay & Evidence recovery, time to first useful evidence, entity-pivot accuracy, and query count. & Recovery of key evidence from closed cases with fewer unsafe or irrelevant steps than baselines. & Criterion evidence for investigation support. \\
Purple-team simulation & Precision, recall, F1, ATT\&CK coverage, correlation quality, and false-positive drivers. & Improved recall without unacceptable false-positive burden against seeded behaviors and benign decoys. & Detection and correlation utility under controlled ground truth. \\
Shadow-mode SOC pilot & Analyst usefulness, triage-time reduction, accepted-pivot rate, trust calibration, and qualitative feedback. & Demonstrated workflow value with no policy bypass and bounded cost. & Naturalistic formative evidence for adoption readiness. \\
Limited production & Latency, availability, cost per hunt, throttling, audit completeness, blocked policy violations, incident contribution. & Stable operation within approved budget, complete auditability, and acceptable reliability. & Context-bound production suitability. \\
\bottomrule
\end{tabularx}
\end{table}

\subsection{Ablation and Negative-Case Analysis}
Causal claims require ablation. The study should compare the full brokered condition with versions that remove schema grounding, approved-template constraints, time-window enforcement, threat-intelligence provenance filters, transparency explanations, and CI/CD gates. Each removal should create a predicted degradation. For example, removing schema grounding should increase invalid-column references. Removing template constraints should increase broad scans and policy violations. Removing threat-intelligence provenance should increase weak indicator matches or expired-indicator use. Removing transparency should reduce analyst comprehension and trust calibration. Negative cases should be classified by root cause: schema failure, prompt interpretation failure, policy failure, stale context, scoring failure, or interface failure.

\subsection{Evidence Maturity}
Table~\ref{tab:evidence} ties claims to evidence maturity. This structure protects the study from overclaiming and gives managers a defensible adoption path.

\begin{table}[!t]
\caption{Evidence Maturity and Claim Boundaries}
\label{tab:evidence}
\centering
\footnotesize
\begin{tabularx}{\columnwidth}{L{0.26\columnwidth}Y}
\toprule
\textbf{Evidence Level} & \textbf{Claim Boundary} \\
\midrule
Level 1: package tests & The artifact is inspectable, reproducible, and technically plausible. \\
Level 2: offline benchmark & The artifact produces safer and more valid output than selected baselines in controlled tasks. \\
Level 3: replay and purple-team & The artifact can recover evidence and detect seeded behaviors under controlled ground truth. \\
Level 4: shadow pilot & Analysts find the artifact useful in supervised real workflow contexts. \\
Level 5: limited production & The artifact operates safely, reliably, and cost-effectively within a defined organizational boundary. \\
\bottomrule
\end{tabularx}
\end{table}

\subsection{Data and Reproducibility Strategy}
Production SOC telemetry is normally unsuitable for public release because it may contain identities, hostnames, email metadata, command lines, URLs, file names, incident details, and other sensitive evidence. The evaluation strategy therefore uses a layered reproducibility approach. The first layer is synthetic Microsoft-like tables that preserve schema and query semantics without exposing production data. The second layer is schema-normalized public traces that allow repeatable KQL evaluation. The third layer is internal shadow-mode evidence that organizations can collect without releasing sensitive telemetry. This approach supports scholarly transparency while respecting security and privacy constraints.

The reproducibility package should include versioned KQL templates, policy files, prompt templates, retrieval schemas, synthetic datasets, expected outputs, linter rules, and execution scripts. It should also include a manifest that records schema versions, prompt versions, model settings, template owners, and benchmark identifiers. Such packaging aligns with DSR transparency because reviewers can inspect the artifact's build logic, evaluation logic, and claim boundaries without requiring live enterprise telemetry.

\begin{table}[!t]
\caption{Reproducibility Artifacts and Review Purpose}
\label{tab:repro}
\centering
\footnotesize
\begin{tabularx}{\columnwidth}{L{0.31\columnwidth}Y}
\toprule
\textbf{Artifact} & \textbf{Review Purpose} \\
\midrule
Synthetic tables & Inspect query semantics without sensitive telemetry. \\
KQL templates & Verify bounded, parameterized, owner-reviewed query design. \\
Policy files & Inspect source allowlists, time limits, row limits, and approvals. \\
Prompt templates & Reproduce AI planning conditions and safety instructions. \\
RAG schemas & Confirm which knowledge sources ground generated output. \\
Benchmark cases & Compare models, prompts, and templates across repeated runs. \\
Manifests and checksums & Establish package integrity and version traceability. \\
\bottomrule
\end{tabularx}
\end{table}

\subsection{Managerial Decision Scorecard}
Managers need a scorecard that combines technical performance with governance and organizational readiness. A technically capable broker should not progress to production if audit completeness is weak, cost behavior is unstable, analysts do not understand the output, or role ownership is unclear. The scorecard in Table~\ref{tab:scorecard} operationalizes this logic. It is intended as a decision aid rather than a universal formula; organizations should adjust thresholds based on risk appetite and regulatory context.

\begin{table}[!t]
\caption{Managerial Decision Scorecard for AI-Enabled Threat-Hunting Adoption}
\label{tab:scorecard}
\centering
\footnotesize
\begin{tabularx}{\textwidth}{L{0.17\textwidth}L{0.26\textwidth}Y Y}
\toprule
\textbf{Dimension} & \textbf{Representative Measures} & \textbf{Acceptable Pilot Signal} & \textbf{Production Concern} \\
\midrule
Technical validity & Schema validity, execution success, invalid-column rate, and expected-field coverage. & High schema validity and stable template behavior. & Repeated hallucinated fields or inconsistent execution. \\
Governance & Policy-bypass blocking, approval compliance, least privilege, and audit completeness. & Complete audit trail and no successful policy bypass. & Missing approvals, broad identities, or incomplete logging. \\
Operational value & Time to first useful evidence, manual pivot reduction, documentation quality, and analyst usefulness. & Analysts accept recommendations and report workflow benefit. & Low usefulness or increased analyst review burden. \\
Cost and reliability & Latency, throttling, timeout rate, model use, and cost per hunt. & Stable operation within approved budgets. & Unbounded scans, unpredictable cloud cost, or high timeout rate. \\
Organizational readiness & Role ownership, training, CI/CD maturity, and escalation paths. & Named owners and complete training for pilot users. & Ambiguous accountability or immature repository controls. \\
Responsible AI & Transparency, uncertainty display, prompt-injection handling, and sensitive-data minimization. & Analysts can explain evidence basis and limitations. & Overreliance, hidden reasoning, or sensitive field exposure. \\
\bottomrule
\end{tabularx}
\end{table}

\section{Contribution to the Body of Knowledge}
\subsection{Scholarly Message, Value, and Impact}
The scholarly message is that AI-assisted operational decision support should be evaluated as a managed socio-technical capability rather than as a model-output problem. KQL supplies the technical instantiation, but the theoretical contribution is the relationship among AI capability, governance control, operational integration, evidence maturity, and managerial value. The value of the study is its explanation of how engineering managers can convert AI from an informal productivity aid into an accountable operating capability by embedding it inside roles, policies, templates, evidence gates, audit trails, and maturity stages. The impact for the IEEE TEM community is a transferable management pattern for high-risk technical functions in which AI produces code-like or evidence-bearing artifacts, but organizational decision rights, validation gates, and accountability remain explicit.

\subsection{Contribution to Engineering Management Theory}
The study contributes to engineering management theory by conceptualizing governed AI adoption as a socio-technical control problem. This extends technology adoption research by emphasizing not only whether users accept AI, but whether the organization can govern AI-generated work products in privileged technical environments, including work products that are read-only or advisory yet still capable of creating organizational risk through exposure, cost, misinterpretation, or uncontrolled diffusion. The design propositions also add to the literature on AI management. P1 and P2 explain how execution control and template ownership reduce the risk of excessive AI agency. P3 explains how intelligence provenance prevents weak external signals from becoming unsupported operational conclusions. P4 links DSR evidence maturity to managerial decision gates. P5 adapts coding-agent research to a security-content engineering context. P6 identifies role-centered accountability as a condition for sustainable AI adoption in SOCs.

\subsection{Contribution to Practice}
The practical contribution is a usable adoption framework. Because information systems value depends on effective use, organizational routines, and capabilities for action under uncertainty \cite{burtonjones2013,pavlou2010}, managers can use the framework to define who owns templates, who approves high-risk hunts, who monitors cost, who validates threat intelligence, who reviews agent-generated changes, and what evidence is required before deployment expands. The artifact also gives practitioners concrete implementation controls: managed identity, read-only adapters, source allowlists, time and row limits, template metadata, RAG schema refresh, prompt-injection tests, audit logs, and CI/CD gates. These controls make the paper actionable for SOC leaders without reducing it to a vendor runbook.

\subsection{Contribution to Cybersecurity Operations Management}
The study extends SOC management research by focusing on evidence governance. Threat hunting depends on uncertain signals, evolving schemas, analyst judgment, and operational constraints. The proposed broker creates a trace from user request to AI plan, retrieved context, generated KQL, broker decision, executed query, returned evidence, analyst disposition, and improvement backlog. This trace improves managerial visibility into where AI helped, where analysts overrode it, where templates failed, and where additional training or controls are needed.

\section{Discussion}
\subsection{Implications for Managers}
The study suggests that managers should not begin AI-enabled SOC adoption by selecting a model. They should begin by defining the decision rights and evidence gates around AI use. The critical questions are: Which data can the system access? Which identities execute queries? Which tables are allowed? What time windows are permitted? Which hunts require approval? What cost budget applies? What evidence must be logged? Who reviews agent-generated changes? What failure conditions require rollback?

Managers should also avoid equating AI fluency with operational readiness. A model can produce readable KQL while still violating schema, cost, privacy, or approval constraints. Therefore, the broker should be evaluated as a control system. Its managerial value comes from reducing unsafe variance in AI-assisted work, making that work inspectable, and giving leaders a defensible basis for scaling or stopping deployment. Its expected impact is improved governance quality, faster but still accountable investigation support, clearer role ownership, and stronger evidence for audit and post-incident review.

\subsection{Implications for Detection Engineering}
Detection engineering becomes a lifecycle discipline under the proposed design. KQL templates, prompts, schemas, threat-intelligence policies, and benchmark cases are versioned assets. AI and coding agents can accelerate maintenance, but they cannot replace content ownership. This framing aligns detection engineering with secure software delivery: every change should be tested, reviewed, documented, and reversible.

\subsection{Implications for Responsible AI}
The artifact operationalizes responsible AI in a concrete engineering context. Excessive agency is controlled by brokered execution and sandboxed coding agents. Sensitive information disclosure is controlled by source policy and projection limits. Prompt injection is addressed by treating telemetry as untrusted evidence rather than instructions. Explainability is addressed by separating evidence from interpretation. Governance is addressed by ERB gates, trace logs, and role accountability. These practices are consistent with AI risk management and LLM application risk guidance \cite{nist_ai_rmf,owasp_llm_top10}.

\section{Threats to Validity}
\subsection{Construct Validity}
Construct validity concerns whether the evaluation measures the intended concepts. KQL validity and execution success measure technical correctness, but they do not fully measure analyst judgment, organizational trust, or managerial value. The evaluation therefore includes analyst usefulness, triage time, evidence quality, audit completeness, and cost visibility. These constructs should be interpreted together rather than as interchangeable performance measures.

\subsection{Internal Validity}
Internal validity concerns whether observed improvements are caused by the broker rather than by better templates, analyst learning, or easier scenarios. The proposed ablation design mitigates this threat by removing one design control at a time. Matched tasks and repeated scenarios also help isolate whether schema grounding, template constraints, policy enforcement, or transparency accounts for observed improvements.

\subsection{External Validity}
External validity concerns whether findings generalize beyond Microsoft-centric SOCs. The architecture is most directly applicable to organizations using Defender XDR, Sentinel, Log Analytics, Sentinel data lake, ADX, and governed DevSecOps workflows. The design principles may transfer to other query ecosystems, but adapters, schemas, permissions, and cost models will differ.

\subsection{Reliability}
Reliability concerns whether results remain stable across repeated runs, model versions, schema updates, and organizational contexts. LLM outputs and coding-agent changes can vary. The study therefore requires prompt versioning, model settings, schema snapshots, benchmark identifiers, and repeated tests. Diachronic reliability should be evaluated after model upgrades, schema refreshes, and template changes.

\section{Limitations and Future Research}
This study has a few limitations. First, the artifact is designed primarily for Microsoft-centric SOCs. The principles may generalize to Splunk, Elastic, or cloud-native SIEMs, but adapters and schemas would differ. Second, the study specifies an evaluation protocol and engineering evidence boundary; it does not claim multi-enterprise production performance. Third, synthetic and schema-normalized public datasets cannot reproduce all tenant-specific telemetry gaps, licensing differences, benign administrative noise, and adversary behavior. Fourth, LLM and coding-agent outputs can vary across model versions, prompts, and tool manifests. Fifth, role-centered governance depends on organizational maturity; organizations without detection-engineering ownership, data-quality processes, or access review discipline should not treat the broker as a substitute for foundational SOC capabilities.

Future research should conduct controlled empirical studies with SOC analysts, compare the broker against manual and unbrokered AI baselines, execute ablation studies, and test the artifact in shadow-mode and limited production settings. Additional work should examine cost governance, trust calibration, analyst skill development, and transferability to non-Microsoft telemetry ecosystems.

The research agenda extends the study along six directions that are relevant to engineering management. First, human-AI work redesign should be studied through controlled analyst tasks, observation, surveys, and interviews to determine when brokered AI reduces cognitive load and when it creates additional review work. Second, governance effectiveness should be tested through ablation studies and policy-bypass experiments that isolate the effect of schema grounding, template constraints, threat-intelligence provenance, and approval gates. Third, cost and reliability management should be evaluated through pilot telemetry, source-routing analysis, latency monitoring, and cost-model simulation. Fourth, detection-engineering maturity should be examined through repository mining, pull-request analysis, and comparison of human-only and coding-agent-supported template maintenance. Fifth, cross-platform transferability should be tested through multi-case DSR replications in Splunk, Elastic, and cloud-native SIEM contexts. Sixth, regulatory defensibility should be evaluated with governance, legal, audit, and compliance stakeholders to identify what evidence package is sufficient for post-incident review and responsible AI assurance.

\section{Conclusion}
This study developed a design science artifact for managing governed AI adoption in security operations. The technical instantiation is an AI-enabled KQL query broker for threat-hunting work, but the scholarly contribution is broader: a management framework for introducing AI-assisted operational decision support into high-risk digital infrastructure while preserving accountability, evidence quality, cost control, privacy, and auditability. KQL is suitable for this purpose because it is widely used for Microsoft security analytics and is read-only in ordinary query use, yet it still exposes meaningful governance concerns around sensitive data, query scope, schema validity, cloud cost, and interpretation quality. The proposed artifact separates AI planning from controlled execution through schema grounding, approved templates, broker validation, read-only adapters, normalized outputs, auditable traces, coding-agent guardrails, and ERB promotion gates. The study contributes to engineering management by showing how a high-risk AI capability can be converted into a controlled socio-technical operating model with explicit roles, evidence gates, quality controls, and evaluation criteria. The central message is that AI-assisted operational decision support should not be managed as unrestricted automation or as a narrow technical tool. It should be managed as a governed engineering capability whose value depends on technical validity, managerial accountability, reproducibility, analyst trust, and evidence-bounded deployment.

\FloatBarrier

\section*{Author Biographies}
\subsection*{Elyson De La Cruz}
is a cybersecurity and information technology executive with over 20 years of experience in enterprise technology leadership, cyber defense, and digital transformation. He is Chief Information Security Officer at Imagine Believe Realize, supporting software and technology initiatives for United States government agencies. His research interests center on decision support in cybersecurity and enterprise systems, especially AI-enabled security analytics, data-driven risk management, machine learning, IoT, and blockchain. He also serves in doctoral advisory and faculty roles at universities in the United States. Dr. De La Cruz is a Distinguished Fellow of the Soft Computing Research Society, a Fellow of the British Computer Society, and a Senior Member of IEEE.

\subsection*{Rishikesh Sahay}
is an Assistant Professor of Cybersecurity at the University of Illinois at Springfield. He is a certified IEC/ISA 62443 Cyber Risk Assessment Specialist. He has worked in the industry as a cybersecurity specialist in Denmark. In industry, he has worked on cyber risk assessment \& mitigation and compliance according to the IEC 62443 standard. He did his Postdoctoral research at the Technical University of Denmark. Prior to that, he did his PhD from Sorbonne University in 2017. His research interests include Cyber risk assessment, threat modeling, autonomic cyber defense, software-defined networking, and LLM applications in cybersecurity. He is a senior member of IEEE.

\subsection*{Md Rasel Al Mamun}
is an Assistant Professor of MIS at the University of Illinois Springfield. Dr. Mamun holds a Bachelor’s and Master of Science in Physics from the University of Dhaka, and a Ph.D. in Business Computer Information Systems from the
University of North Texas. After his Ph.D., Dr. Mamun worked for 1 year as a Postdoctoral Research Associate in Data Analytics at the University of North Texas and for 3 years as a Lecturer in Business Analytics at Old Dominion University. Dr. Mamun’s primary research interests include solving problems related to emerging technology
usage as well as managing associated privacy and security threats.

\end{document}